\documentstyle{article}
\def\om{\omega}
\def\bx{{\bf x}}
\def\bp{{\bf p}}
\def\bu{{\bf u}}
\def\bv{{\bf v}}
\def\by{{\bf y}}
\def\bz{{\bf z}}
\def\bP{{\bf P}}
\def\nn{\nonumber}

\author{
Piret Kuusk \\
Institute of Physics, Riia 142, Tartu EE2400, Estonia\\
piret@fi.tartu.ee\\
\and
J\"uri \"Ord \\
Tartu University, T\"ahe 4, Tartu EE2400, Estonia\\
jyri@hexagon.park.tartu.ee
}
\title{
Kinematics and uncertainty relations of a quantum   
test particle in a curved space-time
\footnote{Based on the talk given in the session ``Quantum Fields
in Curved Space'' at the VIII Marcel Grossmann Conference in Jerusalem, 
Israel, June 1997}
}
\date{\mbox{}}

\begin{document}
\maketitle
\begin{abstract}
A possible model for quantum kinematics of a test particle 
in a curved space-time is proposed. 
Every reasonable neighbourhood $V_e$ of a curved space-time 
can be equipped with a 
nonassociative binary operation called the geodesic multiplication
of space-time points. 
In the case of the Minkowski space-time, left and right translations 
of the geodesic multiplication coincide
and amount to  a rigid shift of the space-time $x \rightarrow x+a$.
In a curved space-time infinitesimal geodesic right translations   
can be used to define the (geodesic) momentum operators.
The commutation relations of position and momentum operators are taken 
as the quantum kinematic algebra.
As an example, detailed calculations are performed for the  
space-time of a weak plane gravitational wave.
The uncertainty relations following from the commutation rules
are derived and their physical meaning is discussed. 
\end{abstract}

\section{Introduction}

The Poincar\'e group -- the symmetry group of the flat space-time $M$ --
and its representations are basic constitutive elements for
relativistic theories. A generic curved space-time $V$ doesn't allow symmetry 
groups and the Poincar\'e group looses its central role. The Lorentz
group can be considered as the symmetry group of flat tangent spaces,
but the status of the Poincar\'e translations is unclear. 

The Poincar\'e translations form an Abelian group and describe rigid 
shifts along straight lines
of a flat space-time,  $x \rightarrow x+a, \> a= const$.
Straight lines are geodesic lines of the Minkowski space. 
In a curved space-time analogous geodesic translations 
can be introduced in a (finite) neighbourhood $V_e$ of $e \in V$
using the concept of geodesic multiplication of points $x,\,y 
\in V_e$. The neighbourhood $V_e$ together with the binary operation
of geodesic multiplication  constitutes an algebraic system
called local geodesic loop. In general, it is  noncommutative and 
nonassociative \cite{Kikkawa, Sabinin, kopjmp}.
As a result we obtain a novel generalization
of Poincar\'e  shifts to the case of a curved space-time (Sec. 2).

In the present paper we investigate some prospects of using
local geodesic loops for constructing a quantum kinematics in the
background of a curved space-time (Sec. 3). 
Let us introduce an action of the position operators $\bx^i$
(on scalar valued functions) as multiplication with the Riemann normal 
coordinates $x^i$.  
We propose to define (geodesic) momentum operators $\bp_k$ via infinitesimal 
right geodesic translations by 
$\bp_k(x)=-i\hbar R^s_k(x)\partial_s$.
The corresponding commutation relations are taken as the quantum 
kinematic algebra \cite{kopcqg}:
$$
{[}{\mathbf x}^i,{\mathbf x}^k{]}=0,\quad
{[}{\mathbf x}^i,{\mathbf p}_k{]}  =i\hbar R^i_k (x),\quad
{[}{\mathbf p}_j,{\mathbf p}_k{]}    =-i\hbar\rho_{jk}^n (x)\,{\mathbf p}_n. 
$$ 
The uncertainty relations which follow from the modified  
${[}{\mathbf x}^i,{\mathbf p}_k{]}$ commutator can put restrictions on
minimal values of coordinates and momenta \cite{Maggiore, Kempf2, Kempf3}. 

As an example, detailed calculations are performed in the case of 
a weak plane gravitational wave background (Sec. 4).

There seems to be also another possibility of using the concept of local 
geodesic loops for constructing  quantum kinematics in the background
of a curved space-time. In the flat space-time quantum field theory, 
one-particle states are 
introduced as representations of the Poincar\'e group and their 
momentum is identified as eigenvalues of the Poincar\'e translation 
operators ${\bf P_\mu}$.  
We could mimic it by defining one-particle  states in a curved
background via representations of geodesic loops. However, 
since the representation theory of general nonassociative structures
is still essentially lacking, we cannot hope a quick progress
along these lines (Sec. 5).
 
\section{Geodesic loops and geodesic translations}

Let us consider a manifold $V$ with a symmetric 
(torsionless) affine connection $\Gamma^\lambda_{\mu \nu} (x) = 
\Gamma^\lambda_{\nu \mu } (x)$.
Let $x, y \in M_e$ be two points of a neighbourhood $V_e$ of 
$e \in V$ such that  geodesic archs between each two points are
uniquely determined. Geodesic multiplication in respect of the
unit element $e$ is defined by the following formula \cite{Kikkawa, 
Sabinin}
\begin{equation}
x \cdot y \equiv L_x y \equiv R_y x = \bigl(\exp_y \circ \tau^e_y
\circ \exp^{-1}_e \bigr) x\,. \label{geodkorr}
\end{equation}
Here exp$_e X$ denotes exponential mapping $X \rightarrow x ,\, 
X \in T_e V, \,  x \in V$, and $\tau^e_y : T_e V \rightarrow T_y V$
is the parallel transport mapping of tangent vectors from
$T_e V$ into $T_y V$ along the unique local geodesic arch joining
the points $e$ and $y$. By $L_x$ and $R_y$ we have defined
the left ($L$) and the right ($R$) translation operators in
analogy with the case of groups. 

From the definition of the
geodesic multiplication (\ref{geodkorr}) it follows that in the 
case of the Minkowski space-time with orthonormal coordinates
$x$, the right and left geodesic translations coincide, 
$R_a^{flat} = L_a^{flat}$ and the geodesic multiplication
$x \rightarrow x \cdot a = a \cdot x $
describes a rigid shift of the space-time, $x \rightarrow x+a$.

Let us introduce the following infinitesimal right translation
matrix:
\begin{equation}
(x \cdot y)^\mu = x^\mu + R^\mu_\nu (x) y^\nu + \ldots ,\qquad
R^\mu_\nu (x)\equiv {\partial (x \cdot y)^\mu \over \partial y^\nu}
\Big|_{y=e}\,. \label{right}
\end{equation}
Matrix $R^\mu_\nu (x)$ can be used to introduce a local frame field
\cite{kopeta, grg}
\begin{equation}
R_\nu (x)\equiv R^\mu_\nu(x){\partial_\mu}\,.
\label{reeperid}
\end{equation}
From Eq. (\ref{right}) it follows that in the unit element $e$
we have 
$R^\mu_\nu (e) = \delta^\mu_\nu  $. 
The commutator of  vector fields $R_\nu (x)$  define the structure functions
$\rho_{\mu \nu}^\sigma (x)$:
\begin{equation}
[R_\mu(x),R_\nu(x)]=\rho_{\mu \nu}^\sigma (x)R_\sigma (x)\,.
\label{AR}
\end{equation}
Let us specify the coordinates $x \in V_e$ to be the Riemann normal 
coordinates, i.e. the equations of geodesics emerging from $e$
are 
\begin{equation}
x^i (t) = X^i t , \qquad X^i \in T_eV \,. \label{riemkoord}
\end{equation}
Now the equations of exponential mapping and parallel transport
which determine the geodesic multiplication (\ref{geodkorr}) 
can be integrated in the neighbourhood of $e$ as power series 
in $x$ \cite{Akivis, kopjmp}  and the following expansion for
structure functions $\rho_{ij}^k(x)$  can be calculated:
\begin{equation}
\rho_{ij}^k(x) = 2 {R^k}_{n[ij]}(e)  x^n  + \ldots\, .
\label{LRapprx}
\end{equation}
Here $R^k_{nij}(e)$ denote the components of the Riemann
curvature tensor at the origin of coordinates $e$. Note that
from the algebraic point of view, they are the
main part of the associator of the geodesic multiplication \cite{Akivis}:
\begin{equation}
\Bigl(\bigl(x(yz)\bigr)^{-1}_L \bigl((xy)z\bigr)\Bigr)^m 
   = {R^m}_{nrs}(e) x^n y^r z^s+\ldots\,.  \label{assots}
\end{equation}
In this sense the emergence of structure functions instead of
structure constants in the commutator (\ref{AR}) is caused 
by the nonassociativity of geodesic multiplication. 

\section{Kinematics of a quantum test particle}

Let us introduce an action of the position operators $\bx^i$
(on scalar valued functions) as multiplication with the Riemann normal 
coordinates $x^i$. Then we have $[\bx^i,\bx^k]=0$. 
We propose 
to define (geodesic) momentum operators $\bp_i$ via infinitesimal 
right geodesic translations (\ref{right}) by 
$\bp_k=-i\hbar R^s_k(x)\partial_s$.
Then,
\begin{equation}
[\bx^i,\bp_k]=[\bx^i,-i\hbar R^s_k(x)\partial_s]
                = i\hbar R^i_k(x).             \label{kanoonk}
\end{equation}
The full (geodesic) kinematic algebra of a quantum test particle in a curved 
space-time now reads \cite{kopcqg}
\begin{equation}
{[}\bx^i,\bx^k{]}=0,\quad
{[}\bx^i,\bp_k{]}  =i\hbar R^i_k (x),\quad
{[}\bp_j,\bp_k{]}    =-i\hbar\rho_{jk}^n (x)\,\bp_n.  \label{geocomrel}
\end{equation}
Note that the modification of commutation relations does not introduce 
any new dimensionful constants.

In a torsionless space-time, an expansion of $R^i_k(x)$ in the Riemann 
normal coordinates ($\Gamma^m_{nr}(e)=0$) can be found by 
using Eq.~(\ref{right}) and the geodesic multiplication 
formula of Akivis \cite{Akivis}
\begin{equation}
(x\alpha)^m = x^m + \alpha^m - {1 \over 2}
\Gamma^m_{nr , s} (e) x^n x^r \alpha^s - {1\over 2}
\Gamma^m_{n (r ,s)} (e) x^n \alpha^r \alpha^s
+ \ldots \,.    \label{akivis}
\end{equation}
We get 
\begin{equation}
{[}\bx^i,  \bp_k{]}=i\hbar\, \left(\delta^i_k -{1\over3}
{J^i}_{kmn}(e)x^m x^n +O(x^3)\right),\label{xpkomm}
\end{equation}
where  
${J^i}_{kmn} $ denotes the Jacobi curvature tensor,
\begin{equation}
{J^i}_{kmn}={1\over2}({R^i}_{mkn}
                           +{R^i}_{nkm}). \label{Jacobi}
\end{equation}
Using Eq.~(\ref{LRapprx}) the expression for $[\bp_i, \bp_j]$
in the Riemann normal coordinates reads
\begin{equation}
{[}\bp_i,\bp_j{]} = -2i\hbar\,\left( {R^k}_{n[ij]}(e)x^n + O(x^2) \right) 
\bp_k \,. \label{bpbp}
\end{equation}
Approximate expressions (\ref{xpkomm}), (\ref{bpbp}) for 
the commutators were presented also  by Kempf \cite{Kempf3}, who introduced
momentum operators as generators of the change of geodesic
coordinates at infinitesimal shift of their origin and used
Synge's world function for calculating commutators.

\section{An example: quantum kinematics in the background of a 
weak plane gravitational wave}

The metric tensor of the space-time of a weak plane
gravitational wave can be given as perturbations around
the Minkowski metric \cite{MTW}:
$$
g_{\mu\nu} = \eta _{\mu\nu} + h_{\mu\nu},
$$
$$
\eta _{\mu\nu} = diag(-1,+1,+1,+1),
$$
In case of a polarized weak plane gravitational wave moving in the direction
of $x^1$ the only non-zero components of $h_{\mu\nu}$ in the TT-gauge  are
\begin{equation}
h_{22} = -h_{33} = A\cos \om (x^0-x^1). \label{TTmeetrika}
\end{equation}
Here $A=const,A\ll 1$, is the wave amplitude, and all subsequent equations
hold in the linear approximation in $A$.

In these coordinates the equation of a geodesic line $x^\mu (t)$
with a tangent vector $X^{\mu}$ at a point $e$ can be easily integrated, 
yielding \cite{kopeta}
$$
x^{\mu}(t) = X^{\mu}t + AU^{\mu}{\sin\om Ct - \om Ct \over \om C^2},
$$
where we have taken $e^\mu = 0$ and denoted
$$
C = X^0 - X^1,
$$
$$
U^0 = U^1 = -{1\over 2}\left((X^2)^2 - (X^3)^2\right) ,
$$
$$
U^2 = - X^2 C \quad ,\quad U^3 = X^3 C .
$$
The coordinates $ X^\mu$ are also the 
Riemann normal coordinates  of a point
$x$ with TT-coordinates $x^\mu \equiv x^\mu (1)$.

Let us choose the point $e$ to be the unit element of the geodesic loop
and let $x,y$ be two points from its neighbourhood.
According to (\ref{geodkorr}), for calculating the product of the 
points $x,y$, the corresponding equations of geodesics 
and of parallel transport of the tangent vector $X^{\mu}$
must be integrated.
From the expression for the geodesic product the matrix
of the right translations and the corresponding structure
functions can be determined \cite{kopeta}.

According to our proposal, canonical momentum operators can be represented 
by $\bp_i = -i R^s_i \partial_s$.
The generalized commutation relations (\ref{geocomrel}) in the background of 
a weak plane gravitational wave read 
\begin{eqnarray*}
[\bx^i,\bx^k] &=& 0,\cr
[\bx^k,\bp_i] &=& i \hbar \left(\delta^k_i + A{\sin \om C - 
\om C \over \om C^3}
(2U^k \partial_i C - \partial_i U^k C)\right), \cr
[\bp_i,\bp_j] &=& {2i \hbar A\over \om C^2}\sin^2{\om C\over 2}
\left(\partial_j U^k \partial_i C - \partial_i U^k \partial_j C\right)\bp_k. 
\end{eqnarray*}
Let us consider more in detail $ [\bx^k,\bp_i]$  commutator.
Near the light-cone emerging from the unit element or in the 
long wavelength approximation, it can be expanded as a series in $\om C$, 
giving a polynomial function on its r.h.s.:
\begin{equation}
[\bx^k,\bp_i] = i\hbar \left( \delta^k_i - a (2 U^k \partial_i C - 
C \partial_i U^k) \right) \,, \label{ligikxp}
\end{equation}
where $a \equiv {1 \over 6} A \om^2 > 0$. Introducing light-cone
coordinates $u = X^0 - X^1$, $v = X^0 + X^1$, $p_u = {1 \over 2}(p_0-p_1)$,
$p_v = {1 \over 2}( p_0 + p_1)$ and denoting $X^2 = y$,
$X^3 = z$, the nonvanishing commutators read
\begin{eqnarray*}
\begin{array}{rclcrcl}
[\bu , \bp_u ] &=& i\hbar,& \qquad & 
        [\bv , \bp_u ] &=& 2i\hbar a (\by^2 - \bz^2), \cr
[\by , \bp_u ] &=& i\hbar a \by \bu ,&\qquad & 
             [\bz , \bp_u ] &=& -i\hbar a \bz \bu, \cr
[\bv , \bp_v ] &=& i\hbar,  &\qquad& \cr
[\bv , \bp_y ] &=& -2i\hbar a\by \bu , &\qquad & 
            [\by , \bp_y ] &=& i\hbar(1-a \bu^2) , \cr
[\bv , \bp_z ] &=& 2i\hbar a\bz \bu , &\qquad &
            [\bz , \bp_z ] &=& i\hbar(1 + a\bu^2). 
\end{array}
\end{eqnarray*}
From these commutators the following uncertainty relations can be derived:
\begin{eqnarray*}
\Delta u \Delta p_u &\geq& {\hbar \over 2}, \cr
\Delta v \Delta p_u &\geq& \hbar a \left( (\Delta y)^2-(\Delta z)^2+
<y>^2-<z>^2 \right), \cr
\Delta y \Delta p_u &\geq& {\hbar a\over 2} \left( \Delta y \Delta u + 
<y><u> \right), \cr 
\Delta z \Delta p_u &\geq& {\hbar a\over 2} \left( \Delta z \Delta u + 
<z><u> \right), \cr
\Delta v \Delta p_v &\geq& {\hbar \over 2} , \nn\cr
\Delta v \Delta p_y &\geq& \hbar a \left( \Delta y \Delta u + 
<y><u> \right), \cr
\Delta y \Delta p_y &\geq& {\hbar \over 2} \left( 1 - a((\Delta u)^2 +
<u>^2) \right), \cr
\Delta v \Delta p_z &\geq& \hbar a \left( \Delta z \Delta u + 
<z><u> \right), \cr
\Delta z \Delta p_z &\geq& {\hbar \over 2} \left( 1 + a((\Delta u)^2 + 
<u>^2) \right). 
\end{eqnarray*}
Consider a state which has $<u>=0$, $<y>=0$ or $<z>=0$. An example of
such a state is a test particle moving along with the wave front.
The uncertainties for $p_u$ then read
\begin{equation}
 \Delta p_u \geq {\hbar \over 2\Delta u}, \quad 
 \Delta p_u \geq {\hbar a \over 2} \Delta u.
\label{pumaaram}
\end{equation}
The uncertainty relations (\ref{pumaaram}) can be combined linearly to give  
\begin{equation}
\Delta p_u \geq {\hbar \over 2}\left( {1 \over \Delta u} + D a\Delta u \right),
\end{equation}
where $D$ denotes an undefined constant. This relation entails
a minimal uncertainty for $p_u$ \cite{Kempf2}
\begin{equation}
(\Delta p_u )_{min} =  \hbar \sqrt{Da}.
\end{equation}
The uncertainty relations for the transverse components of the
momentum operator are more complex, those for $\Delta p_v$ 
remain unmodified.

\section{Some remarks about the quantum field theory}

We have presented possible kinematics of a relativistic quantum
test particle in a curved space-time. 
In a flat space-time, the full first-quantized theory of a
relativistic particle is plagued by negative probabilities and
must be replaced by a second-quantized quantum field theory.

We may try to mimic the flat space-time quantum field theory
by replacing the momentum operators $\bP_k =-i\hbar \partial_k $
with the generalized momentum operators defined via the infinitesimal 
right geodesic translation operators
$\bp_k =-i\hbar R^s_k(x)\partial_s$. In a flat space-time,
components of the momentum operator commute, $[\bP_j, \bP_k]=0$.
In a curved background we have instead, 
\begin{equation}
{[}\bp_j,\bp_k{]}  = -i\hbar\rho_{jk}^n (x)\,\bp_n. 
\end{equation}
For instance, in the case of a weak plane gravitational wave
background, in the approximations considered in the last section 
we have two nonvanishing commutators:
\begin{eqnarray*}
[\bp_y, \bp_u]& =& {i \hbar \over 2} A \om (2 \by \bp_v - \bu \bp_y),\cr
[\bp_z, \bp_u]&=& -{i \hbar\over 2}A \om (2 \bz \bp_v + \bu \bp_z).
\end{eqnarray*} 
It follows that the time-like component of the momentum $\bp_t = \bp_u+\bp_v$ 
doesn't commute with the transverse components of the momentum,
$[\bp_t , \bp_y] \not= 0$, $[\bp_t , \bp_z] \not= 0$. If $\bp_t$
could be interpreted as the energy operator (hamiltonian) and if 
an analog of the Noether theorem for geodesic loops could be established, 
this means that the transverse momentum of the quantum field is not
conserved in the background of a weak plane gravitational wave.
There has been some progress in establishing generalized
conservation laws in the case of Moufang loops \cite{lps}, but nothing
can be said in the case of more general geodesic loops.
The representation theory of general nonassociative structures
is also essentially lacking and we cannot introduce one- and many-particle
states as suitable representations of geodesic loops. It seems that
some novel mathematical ideas and developments are needed for
continuing our investigations in this direction. 

\bigskip\noindent
{\Large \bf Acknowledgements}

\nopagebreak
\medskip\noindent
The authors are grateful to Dr.\ Eugen Paal for the idea to use 
geodesic loops in the theory of gravitation and enlightening 
discussions on nonassociative algebraic systems.   
This work was supported by the Estonian Science Foundation under
Grants No. 2279 and 2403.



\begin{thebibliography}{xx}

\bibitem{Kikkawa}
\newblock{M. Kikkawa,   J. Hiroshima Univ. Ser A--1. Math.
{\bf 28}, 199 (1964).}

\bibitem{Sabinin}
\newblock{L.V. Sabinin, Dokl. Akad. Nauk SSSR {\bf 233}, 800
(1977) (in Russian);
Trans. Inst. Phys. Estonian Acad. Sci. {\bf 66}, 24 (1990).}

\bibitem{kopjmp}
\newblock{P. Kuusk, J. \"Ord, and E. Paal, J. Math. Phys. {\bf 35}, 321
(1994).}

\bibitem{kopcqg}
\newblock{P. Kuusk, J. \"Ord, and E. Paal, Class. Quant. Grav. 
(to be published).}

\bibitem{Maggiore}
\newblock{M. Maggiore, Phys. Lett. {\bf B304}, 65; {\bf B319},
83 (1993).}

\bibitem{Kempf2}
\newblock{A. Kempf, J. Math. Phys. {\bf 35}, 4483 (1994);
Phys. Rev. {\bf D52}, 1108 (1995), {\bf D54}, 5174 (1996).}

\bibitem{Kempf3}
\newblock{A. Kempf, Preprints hep-th/9405067 and hep-th/9603115
(unpublished).}

\bibitem{kopeta}
\newblock{P. Kuusk, J. \"Ord, and E. Paal,  Proc. Estonian Acad. Sci. 
Phys. Math. {\bf 44}, 437 (1995). }

\bibitem{grg}
\newblock{P. Kuusk, and E. Paal, Gen. Rel. Gravit. {\bf 28} 991 
(1996).}

\bibitem{Akivis}
\newblock{M.A. Akivis, Sibirski Mat. J. {\bf 19}, 243 (1978) (in Russian).}

\bibitem{MTW}
\newblock{C.W. Misner, K.S. Thorne, and J.A. Wheeler, {\sl Gravitation}
(Freeman, San Francisco, 1973).}

\bibitem{lps}
\newblock{J. L\~ohmus, E. Paal, and L. Sorgsepp,   Contemp. Math. 
{\bf 184} 281 (1995).} 

\end{thebibliography}
\end{document}